# Quantum-level control in a III-V-based ferromagnetic-semiconductor heterostructure with a GaMnAs quantum well and double barriers

Shinobu Ohya[a]

*Department of Electrical Engineering and Information Systems, The University of Tokyo, 7-3-1 Hongo, Bunkyo-ku, Tokyo 113-8656, Japan, and PRESTO Japan Science and Technology Agency, 4-1-8 Honcho, Kawaguchi, Saitama 332-0012, Japan*

Iriya Muneta and Masaaki Tanaka[b]

*Department of Electrical Engineering and Information Systems, The University of Tokyo, 7-3-1 Hongo, Bunkyo-ku, Tokyo 113-8656, Japan*

We investigate the spin-dependent tunneling properties in a three-terminal III-V-based ferromagnetic-semiconductor heterostructure with a 2.5-nm-thick GaMnAs quantum well (QW) and double barriers. We successfully control the quantum levels and modulate the spin-dependent current with varying the voltage of the electrode connected to the GaMnAs QW. Our results will open up a new possibility for realizing three-terminal spin resonant-tunneling devices.

a) Electronic mail: ohya@cryst.t.u-tokyo.ac.jp
b) Electronic mail: masaaki@ee.t.u-tokyo.ac.jp

III-V-based ferromagnetic-semiconductor GaMnAs heterostructures are one of the most ideal systems for future semiconductor-based spintronic devices. Especially, the high coherency of the valence-band holes in these structures is a very promising feature for devices using the quantum-size effect. Recently, we have clearly observed the resonant tunneling effect and the increase of tunnel magnetoresistance (TMR) induced by resonant tunneling in the GaMnAs quantum-well (QW) double-barrier (DB) heterostructures (QWDBs).[1] One of the next important goals is the realization of "three-terminal" GaMnAs QWDB devices such as spin resonant-tunneling transistors,[2] where the spin-dependent quantum levels can be externally controlled by applying the voltage to the electrode connected to the GaMnAs QW. In comparison with the metal-based magnetic tunnel transistors,[3,4] we can more effectively design the heterostructure for controlling the spin-polarized coherent holes utilizing the well-established band-engineering technique of semiconductors. Furthermore, the metallic feature of GaMnAs allows us to make a good contact to the GaMnAs QW layer, thus more efficient control of the quantum levels is possible than that in usual semiconductor-based three-terminal quantum heterostructures containing a QW electrode.[5] Here, we demonstrate the quantum-level control in a three-terminal GaMnAs QWDB heterostructure and show the successful modulation of the spin-dependent current in this device.

Figure 1 illustrates the schematic cross-sectional structure of the device investigated here. The GaMnAs QWDB device was grown by molecular beam epitaxy and consists of, from the surface side, $Ga_{0.94}Mn_{0.06}As$(10 nm)/ GaAs (1 nm)/ $Al_{0.94}Mn_{0.06}As$(4 nm)/ GaAs(2 nm)/ $Ga_{0.94}Mn_{0.06}As$ QW(2.5 nm)/ GaAs(1 nm)/ AlAs(4 nm)/ GaAs:Be (100 nm) on a $p^+$GaAs(001) substrate. The Be concentration of the Be-doped GaAs (GaAs:Be) layer was $2\times10^{18}$ cm$^{-3}$. The thin GaAs spacer layers were inserted to smooth the surface. The GaAs:Be, AlAs and the lowest GaAs spacer layers were grown at high temperatures of 600, 550, and 600 ºC, respectively. The GaMnAs QW, GaAs/ AlMnAs/ GaAs, and the top

GaMnAs layers were grown at low temperatures of 240, 200, 225ºC, respectively. In this structure, we used AlMnAs as the upper tunnel barrier. We confirmed that AlMnAs acts as a paramagnetic tunnel barrier against GaMnAs with a barrier height of 110 meV.[6] The Curie temperature ($T_C$) of the GaMnAs layers was estimated to be ~60 K by magnetization measurements. After the growth, we fabricated a ring-shaped electrode (named QW) whose area is 36-times larger than that of the central electrode (named TOP) on the sample, and the region sandwiched between these electrodes was carefully etched to the depth of the AlMnAs barrier layer. Due to the large difference in area size between these electrodes, the energy potential of the GaMnAs QW can be effectively controlled by the voltage of QW ($V_{QW}$). In the following spin-dependent transport measurements, the TOP electrode is grounded, and we focus on the current $I$ through TOP to the ground when changing $V_{QW}$ and the voltage $V$ of the substrate (named SUB). Note that the sign of $V$ is the opposite to that defined in our previous papers,[1,7] so the resonant levels are detected in the $V>0$ region when $V_{QW}$ is not applied.

Figure 2(a) shows the $dI/dV$-$V$ curves in parallel magnetization when $V_{QW}$ is varied from -0.05 to +0.05 V (from bottom to top) with the voltage step of 0.01 V at 3.6 K. In every curve, oscillations due to resonant tunneling of the heavy-hole first state (HH1), light-hole first state (LH1), and heavy-hole second state (HH2) are observed. The valleys of these oscillations linearly shift to the higher voltages with increasing $V_{QW}$ as shown by the black dotted lines. Figure 2(b) shows the schematic valence-band diagrams of our GaMnAs QWDB heterostructure in terms of hole energy when $V_{QW}$ is positive (upper graph), zero (middle graph), and negative (bottom graph).[8] Black, red and blue lines are the valence band at the Γ point ($E_v$), resonant levels, and the chemical potential $\mu$ of the GaAs:Be electrode. As can be seen in these pictures, when $V_{QW}$ is increased from negative to positive, $V$ for detecting these resonant levels is increased. Thus, the resonant peaks are shifted to higher voltages with increasing $V_{QW}$, which indicates that the quantum levels are successfully controlled by changing $V_{QW}$. In Fig. 2(a), we can also see other oscillations indicated by lines A and B, which will be discussed later.

Figure 3 shows the mappings of (a) $dI/dV$ and (b) $d^2I/dV^2$ as functions of $V_{QW}$ and $V$ in parallel magnetization at 3.6 K. Here, the resonant levels of HH1, LH1, and HH2 are traced by the black dotted curves. The shift in $V$ of these resonant levels is almost saturated when $|V_{QW}|$ gets larger than ~0.2 V. This reason is explained as follows. $V_{QW}$, corresponding to the voltage drop between TOP and QW through the current path indicated by $I''$ in Fig. 1, is mainly consumed at the AlMnAs barrier beneath the TOP electrode and in the GaMnAs QW plane between the TOP and QW electrodes.[9] With increasing $|V_{QW}|$, the tunnel resistance of AlMnAs becomes much smaller due to its low barrier height (~110 meV),[6] while the resistance of the GaMnAs QW plane does not depend on $V_{QW}$. As a result, $V_{QW}$ is consumed mostly in the GaMnAs QW plane when $|V_{QW}|$ is large, thus the shift of the resonant peaks is saturated. We note that the resonant levels are detected when the energy region occupied by holes in the GaAs:Be electrode crosses the resonant levels in the GaMnAs QW regardless of the current direction. Thus, the resonant levels are detected both in $V>0$ and $V<0$ regions. In our studies including other three-terminal GaMnAs QWDB devices (not shown), we found that the shift of the resonant levels tends to be larger when the $T_C$ of the GaMnAs QW is higher, which means that the metallic nature of the GaMnAs QW is important for improving the controllability of the quantum levels.

In Fig. 3(a), there is a deep valley traced with the white dotted curve A passing through the origin ($V=V_{QW}=0$). This valley corresponds to the line A in Fig.2 (a) and the curve A in Fig. 3(b). This valley is considered to be the bias condition where the chemical potentials of TOP and QW are the same. If the potential of the GaMnAs QW were perfectly controlled by $V_{QW}$, these valleys would not be observed. This result means that the potential of the GaMnAs QW is influenced also by $V$ as well as by $V_{QW}$. We can see a small valley traced with the curve B in Fig. 3(a), which can be more clearly seen in Fig. 2(a) (line B) and Fig. 3(b) (white dotted curve B). They correspond to the bias condition where the chemical potentials between QW and SUB are the same.

In Fig. 3(a), there are valleys traced with the white lines C moving almost linearly with changing $V_{QW}$, which overlap the above-mentioned resonant levels when $|V_{QW}|$ is smaller than ~0.05 V. These valleys correspond to the resonant levels formed in the area of the GaMnAs QW plane beneath the ring-shaped QW electrode (See Fig. 1), where the potential is almost ideally controlled by $V_{QW}$ due to its proximity to the QW electrode. Since the GaMnAs QW is metallic, (probably non-ballistic component of) the current $I$ collected at the TOP electrode is expected to be widely spread in the GaMnAs QW plane. Thus, a part of this current is detected in the measurements of $I$. These valleys were more clearly seen in the $dI_{QW}/dV$ mapping as functions of $V_{QW}$ and $V$ (not shown), where $I_{QW}$ is the current going out through the QW electrode.

Figure 4 shows the $V$ dependence of the magneto-current (MC) ratio defined by $(I_P - I_{AP})/I_{AP}$ with various $V_{QW}$ from 0 to -0.14 V at 3.6 K, where $I_P$ ($I_{AP}$) represents $I$ in parallel (anti-parallel) magnetization. (For the measurements of the spin-dependent current, see our previous paper.[1]) To control the magnetization alignment of the GaMnAs layers, the magnetic field was applied along the [100] axis in plane. These MC vs. $V$ data were mathematically derived from the $I$-$V$ data in parallel and anti-parallel magnetizations at zero magnetic field. With changing $V_{QW}$ from 0 to -0.14 V, the MC peak at LH1 is decreased, while the MC peak at HH2 is increased. This opposite behavior is due to the increasing current from TOP to QW identified by $I''$ in Fig. 1 with changing $V_{QW}$ from 0 to -0.14 V. Here, we define this current as $I''_P$ ($I''_{AP}$) and the direct current transferred between TOP and SUB identified by $I'$ in Fig. 1 as $I'_P$ ($I'_{AP}$) in parallel (anti-parallel) magnetization, then MC is expressed as $[(I'_P - I''_P) - (I'_{AP} - I''_{AP})]/(I'_{AP} - I''_{AP})$. Here, we take the MC-$V$ curve with $V_{QW}$=-0.03 (light blue curve) as an example for explaining the MC-$V$ behavior shown in Fig. 4. When $V \ll 0$, MC is dominated by $I'_P$ and $I'_{AP}$, thus the typical $V$ dependence of MC appears, where MC is monotonically decreased with increasing $|V|$. When $V$ is around zero, $I'_P$ and $I'_{AP}$ are comparable to $I''_P$ and $I''_{AP}$, respectively. Thus, the denominator in the definition of MC becomes close to zero, and MC becomes infinity. When $V$ is increased more, the sign of the denominator is changed, thus MC becomes negative infinity. In the $V \gg 0$ region, $I'_P$ and $I'_{AP}$ are dominant again, thus MC becomes positive and monotonically decreases to zero. Although the LH1 peak near the origin is more largely affected by $I''_P$ and $I''_{AP}$ and MC is decreased when $V_{QW}$ is changed from 0 to -0.04 V, the MC peak at HH2 is increased due to the smaller influence of $I''_P$ and $I''_{AP}$. The MC increase at HH2 in the range of $V_{QW}$ from 0 to -0.1 V is mainly because the energy of HH2 becomes close to the chemical potential of TOP where a high spin polarization is expected. In this $V_{QW}$ range, the bias voltage of the MC peak at HH2 moves toward the negative direction with changing $V_{QW}$ from 0 to -0.1 V, following the HH2 peak's shift as shown in Fig. 2(a), 3(a), and (b). This means that we successfully controlled the spin-dependent current by electrically modulating the quantum levels in the GaMnAs QW. When $V_{QW}$ is changed from -0.1 to -0.14 V, the MC peak at HH2 moves toward the positive direction. In Fig. 3(a) and (b), we can see that the resonant level of HH2 is merged with the valley A when $V_{QW}$ is ~-0.1 V and the valley A becomes dominant when $V_{QW}$ <-0.1 V. Thus, the TMR increase observed when $V_{QW}$ < -0.1 V is attributed to the valley A.

In the present device, the power gain has not been obtained due to the large tunneling current $I''$ between TOP and QW. If we can insert a thick barrier with a low barrier height between TOP and QW without suppressing MC, the spin-transistor operation with a power gain is expected to be obtained, which will lead to quantum spin devices, such as spin resonant-tunneling transistors and magnetic monostable-bistable transition logic elements.[10]

In summary, we have fabricated the three-terminal GaMnAs QWDB device. We controlled the quantum levels of the 2.5-nm thick GaMnAs QW and modulated the spin-dependent current with varying $V_{QW}$, which is largely attributed to the metallic feature of the GaMnAs QW. The MC ratio was increased by decreasing the potential of the GaMnAs QW in terms of hole energy by applying the negative bias voltage to the QW electrode.

This work was partly supported by Grant-in-Aids for Scientific Research No. 18106007, No. 19048018, and No. 20686002, the Special Coordination Programs for Promoting Science and Technology, R&D for Next-generation Information Technology by MEXT, and PRESTO of JST.

**References**

[1]  S. Ohya, P. N. Hai, Y. Mizuno, and M. Tanaka, Phys. Rev. B **75**, 155328 (2007).

[2]  Y. Mizuno, S. Ohya, P. N. Hai, and M. Tanaka, Appl. Phys. Lett. **90**, 162505 (2007).

[3]  K. Mizushima, T. Kinno, T. Yamauchi, and K. Tanaka, IEEE Trans. Magn. **33**, 3500 (1997).

[4]  S. van Dijken, X. Jiang, and S. S. P. Parkin, Appl. Phys. Lett. **83**, 951 (2003).

[5]  Lepsa, M.I., van de Roer, Th.G., Kwaspen, J.J.M., Smalbrugge, E., van der Vleuten, W., Kaufmann, L.M.F. 1997 International Semiconductor Conference 20th Edition. CAS'97 Proceedings (Cat. No. 97TH8261) **1**, 139 (1997).

[6]  S. Ohya, I. Muneta, P. N. Hai, and M. Tanaka, Appl. phys. Lett. **95**, 242503 (2009), arXiv:0912.3045.

[7]  S. Ohya, I. Muneta, P. N. Hai, and M. Tanaka, submitted.

[8]  In Ref. 7, we precisely estimated the energy difference between $E_v$ and the Fermi level and found that it depends on the GaMnAs QW thickness.  Here, we neglect this effect for simplicity.

[9]  There is a small current path through the underlying GaAs:Be layer and the AlAs barrier as indicated by dotted path in Fig. 1, but we neglected this path.  We also neglected the resistance of the AlMnAs barrier just beneath the ring-shaped QW electrode, because it is much smaller than that beneath the TOP electrode due to the large difference in area size between the TOP and QW electrodes.

[10]  C. Ertler and J. Fabian, Phys. Rev. B **75**, 195323 (2007).

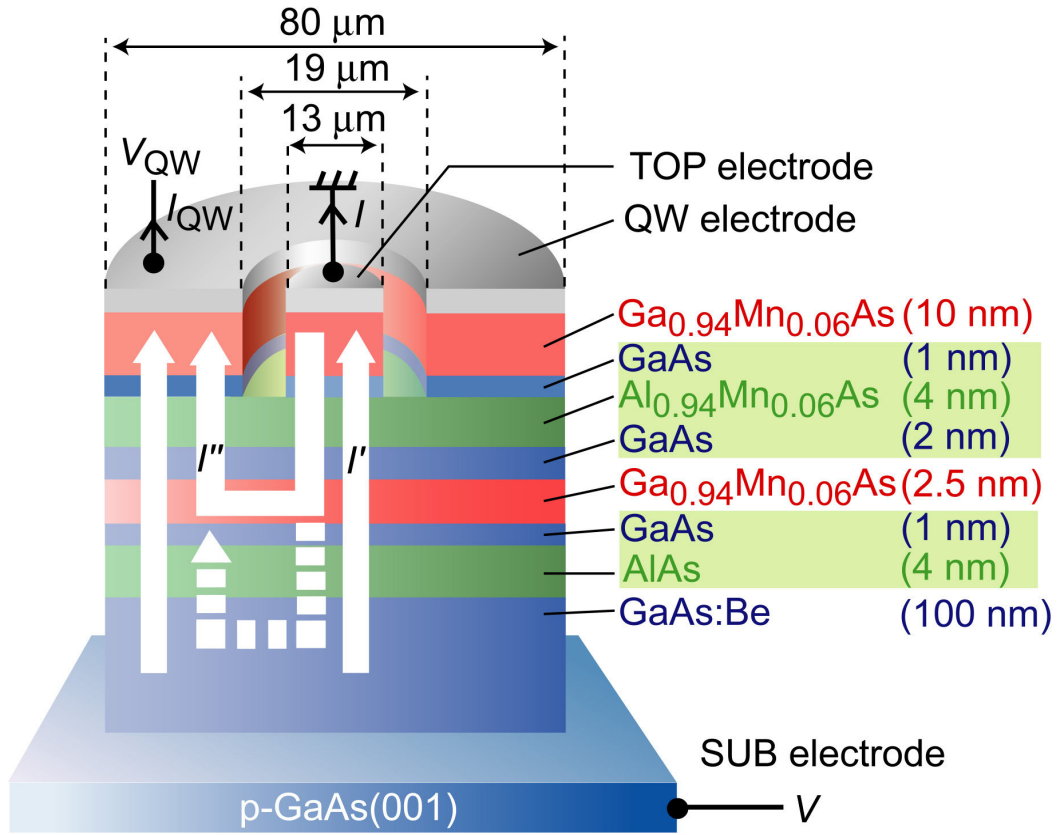

FIG. 1. Schematic cross-sectional structure of the three-terminal device investigated here. The GaMnAs QWDB heterostructure comprises, from the surface side, $Ga_{0.94}Mn_{0.06}As$(10 nm)/ GaAs (1 nm)/ $Al_{0.94}Mn_{0.06}As$(4 nm)/ GaAs(2 nm)/ $Ga_{0.94}Mn_{0.06}As$ QW(2.5 nm)/ GaAs(1 nm)/ AlAs(4 nm)/ GaAs:Be (100 nm) on a p$^+$GaAs(001) substrate. $I'$ represents the current path directly flowing from SUB to TOP. $I''$ is the current path flowing from the TOP electrode to the QW electrode through the GaMnAs QW plane. Dotted path is the one through the GaAs:Be layer. Note that there is also a current path directly from SUB to QW.

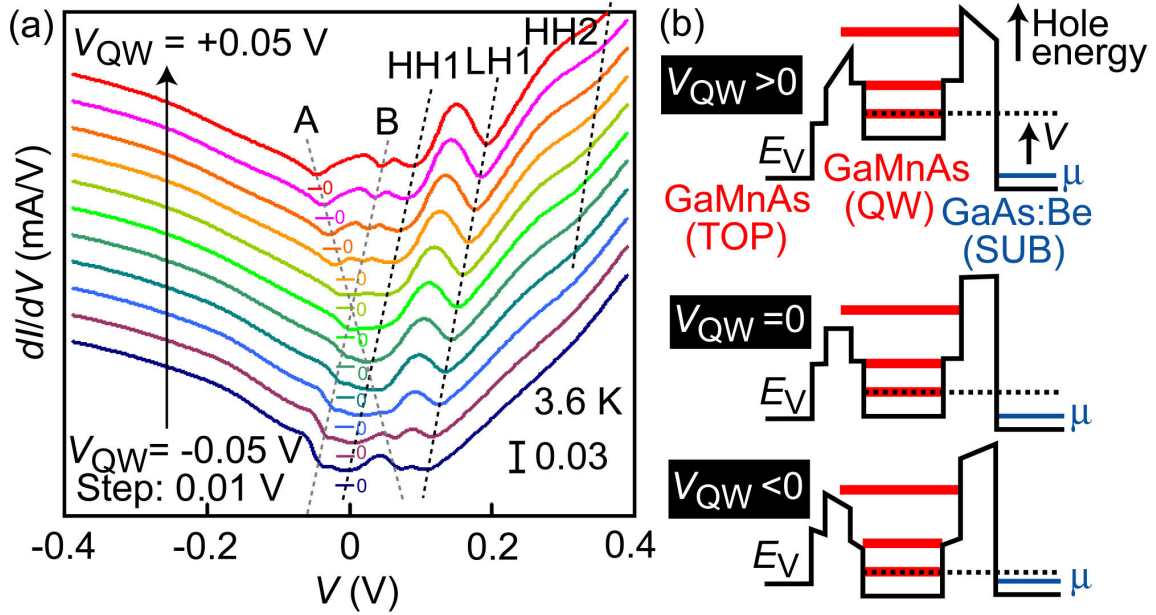

FIG. 2. (a) *dI/dV-V* curves in parallel magnetization when $V_{QW}$ is varied from -0.05 to +0.05 V (from bottom to top) with the voltage step of 0.01 V at 3.6 K. (b) Schematic valence-band diagrams of our GaMnAs QWDB heterostructure in terms of hole energy when $V_{QW}$ is positive (upper graph), zero (middle graph), and negative (bottom). Black, red and blue lines are the valence band at the Γ point ($E_V$), resonant levels, and the chemical potential $\mu$ of the GaAs:Be electrode.

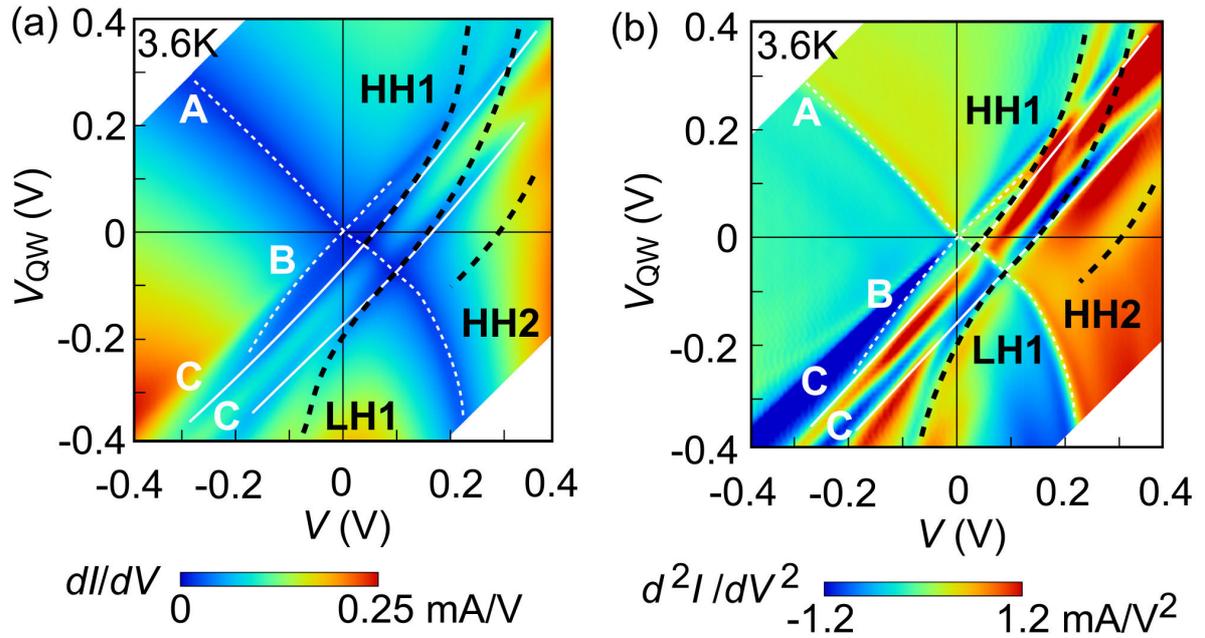

FIG. 3. (a) $dI/dV$ and (b) $d^2I/dV^2$ mappings as functions of $V_{QW}$ and $V$ in parallel magnetization at 3.6 K. Here, the resonant levels of HH1, LH1, and HH2 are traced by the black dotted curves. For the explanation of other curves A-C, see the text.

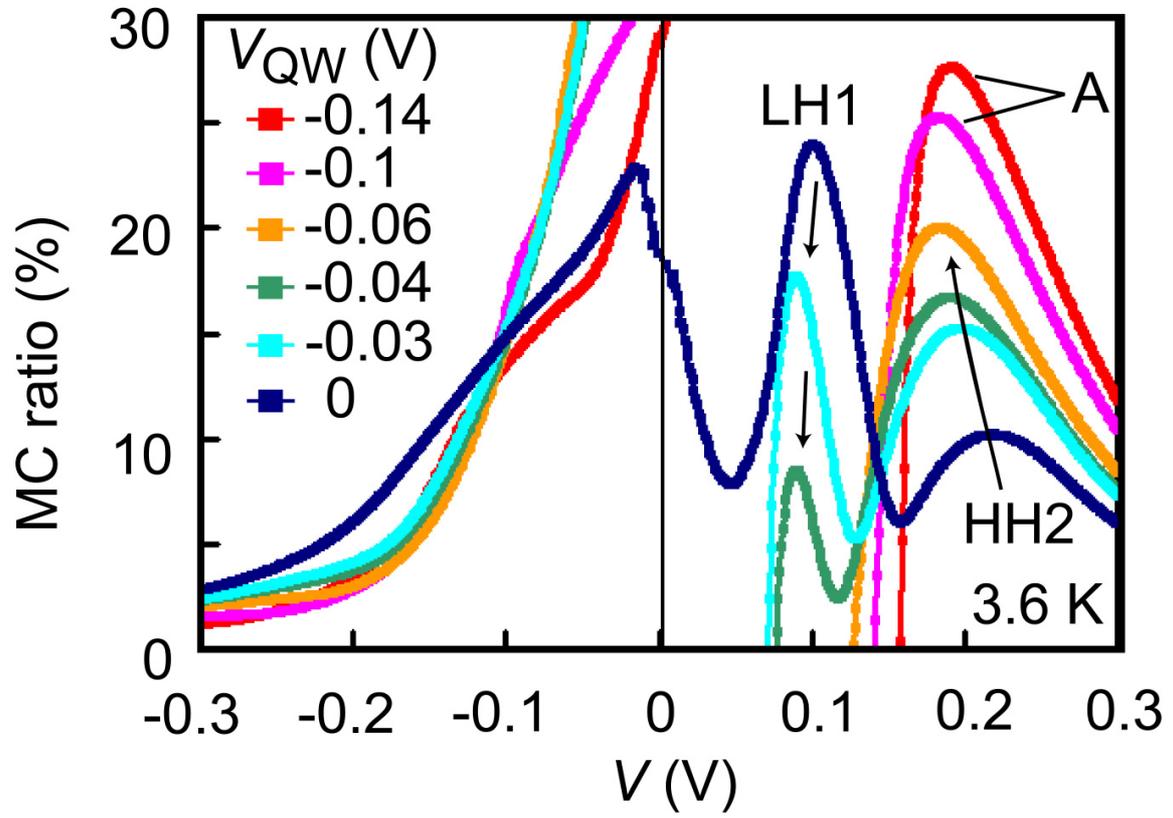

FIG. 4. $V$ dependence of MC defined by $(I_P - I_{AP})/I_{AP}$ at 3.6 K with various $V_{QW}$ from 0 to -0.14 V, where $I_P$ ($I_{AP}$) represents $I$ in parallel (anti-parallel) magnetization.